\documentclass[11pt,a4paper]{article}
\usepackage{jheppub}
\usepackage{amsmath,amsthm,amssymb,graphics,epsfig,color,mathrsfs,cases,colortbl,framed}
\usepackage{ulem}
\usepackage{graphicx}
\usepackage{subfigure}
\usepackage{booktabs}    
\usepackage{multirow}
\usepackage{tabularx}
\usepackage{longtable}
\usepackage{array}
\usepackage{nicematrix}
\usepackage{makecell}
\usepackage{graphicx}
\usepackage{bm}

\title{Fermionic Kaluza-Klein mode mixing in braneworlds}
\author{ Chun-E Fu,}
\author[1]{Wen-Xuan Ma}

\affiliation{
Institute of Modern Physics, Department of Modern Physics and Astronomy, School of Physics, Xi'an Jiaotong University, Xi'an, Shaanxi, China
}

\emailAdd{fuche13@mail.xjtu.edu.cn}
\emailAdd{m\_fuminoki@foxmail.com}

\footnotetext[1]{Corresponding author.}

\abstract{
We investigate fermionic Kaluza-Klein (KK) mode mixing in thick braneworld models subjected to generic background perturbations. {Conventionally, isolated static backgrounds are completely described by a Schr\"{o}dinger-like formulation, which yields an unperturbed orthogonal basis of KK eigenstates. However, generic perturbations possess a non-trivial spatial profile along the extra dimension. When the full interacting Dirac operator is expanded in this original basis, the spatial variation inevitably yields non-vanishing overlap integrals between distinct KK levels, thereby inducing off-diagonal couplings in the 4D effective mass matrix. Consequently, the original eigenstates are no longer exact physical eigenmodes of the perturbed system. } To rigorously preserve the underlying 5D chiral structure and resolve the true physical states, we employ an exact Singular Value Decomposition (SVD) of the full, off-diagonal Dirac mass matrix. Our exact analysis reveals that this mode mixing introduces small but highly structured corrections to the mass eigenvalues. Specifically, parity-odd perturbation operators strictly induce same-parity mixing that preserves the macroscopic $Z_2$ spatial symmetry, whereas parity-even operators trigger cross-parity mixing that shatters the $Z_2$ symmetry, resulting in severe spatial polarization of the KK probability densities. {Phenomenologically, such polarization shifts the wave functions toward the brane, turning probability zeros into non-zero values, which directly illuminates previously ``dark'' KK modes.}
 
}


\begin{document}

\maketitle
\flushbottom

\section{Introduction}
\label{sec:introduction}

Braneworld scenarios with large or warped extra dimensions have provided a compelling framework for addressing fundamental problems in high-energy physics, most notably the gauge hierarchy problem \cite{ArkaniHamed1998, Randall1999, Randall1999a}. Within these geometric setups, the dynamics of bulk fermions have garnered significant attention. In thick braneworld models, the standard formulation relies on some background scalar fields to localize these fermionic fields (through Yukawa couplings) \cite{Rubakov1983,RandjbarDaemi2000, Ichinose2002, Melfo2006,Liu2007,Gogberashvili2007,Liu2009a,Castro2011,Guo2014fba,Dantas2015,Li2017,Paul2017,Moreira2024xyz}. 

The traditional procedure to determine the Kaluza-Klein (KK) mass spectrum and extra-dimensional wavefunctions relies on constructing a Hermitian 5D Dirac operator, denoted as $\hat{D}_0$. This Hermitian operator allows the first-order Dirac equations to seamlessly decouple into two isolated second-order Schr\"{o}dinger-like equations for the left- and right-handed components. The eigenfunctions of these decoupled equations naturally serve as the complete orthogonal basis for the KK decomposition. Crucially, the Hermiticity of $\hat{D}_0$ mathematically guarantee that the resulting mass matrix is strictly diagonal, dictating that absolutely no left-right mixed couplings exist between distinct KK levels.

However, realistic braneworld dynamics are intrinsically more complex than such highly idealized models \cite{DeWolfe2000, Csaki2000fc,Deng2025fbr}. The presence of bulk scalar field fluctuations (e.g., non-minimal dilaton couplings) or geometric metric backreactions inevitably introduces non-trivial distortions to the bulk Dirac operator. If one were to directly absorb these generic perturbations into a newly defined total 5D Dirac operator, the resulting Schr\"{o}dinger-like equations would become prohibitively complicated and analytically intractable. This difficulty arises precisely because generic dynamical fluctuations lack the highly symmetric spatial structure—such as the strict $Z_2$ parity.

Instead, a more rigorous and tractable approach is to treat the explicit perturbation operator independently. By superimposing its action onto the unperturbed system and evaluating it within the original orthogonal reference basis, the generic geometric profile of the perturbation generates non-zero off-diagonal matrix elements. This explicitly destroys the diagonality of the total mass matrix and induces dynamic \textit{left-right mixed couplings} between distinct KK levels. 

Since the late 1990s, the concept of left-right chiral mixing within extra-dimensional frameworks has attracted considerable attention. Historically, its most celebrated phenomenological application has been in the neutrino sector \cite{Dienes1998sb, Grossman1999ra,Barbieri2000gf, Lukas2000hd, Huber2003gh, Fong2011yg,Stenico2018zxy, Dvali2023uxn,Panda2024psj}, where it was proposed as an elegant mechanism to generate tiny masses without invoking the grand unification scales typically required by the canonical seesaw mechanism \cite{Minkowski1977, Yanagida1979, GellMann1979, Mohapatra1980}. Consequently, a rich literature has developed around this specific application (see, e.g., Ref.~\cite{deGiorgi2025ydn} for a recent comprehensive review).

Diverging from typical phenomenological model-building---where mixed couplings are often introduced \textit{ad hoc} as free parameters to accommodate experimental data---the present work is devoted to a theoretical investigation of the universal mechanics governing this KK mode mixing. To rigorously resolve this extensive chiral entanglement, we construct the complete non-diagonal first-order Dirac mass matrix and perform an exact Singular Value Decomposition (SVD). By systematically analyzing this non-perturbative mixing under non-minimal dilaton couplings and geometric fluctuations, we uncover several profound physical regularities that uniquely govern the mixed braneworld fermions:
\begin{itemize}
    \item \textbf{Constraints on the physical mass spectra:} The dynamical mode mixing introduces minor corrections to the mass eigenvalues, which remains perfectly consistent with our assumption of small background perturbations. Despite these shifts, the zero mode is strictly protected. Because only the left-handed component possesses a localizable zero mode, it lacks a right-handed counterpart to mix with, thereby ensuring its mass remains exactly zero.
        
 \item \textbf{Parity selection rules and dynamical spatial polarization:} The KK mode mixing structure and the intrinsic geometric reshaping of physical wavefunctions are strictly dictated by the parity of the perturbation operators. Specifically, parity-odd perturbation operators yield a block-diagonal mass matrix, driving pure same-parity state mixing that perfectly preserves the macroscopic $Z_2$ spatial symmetry via amplitude modulations. In stark contrast, parity-even operators trigger severe cross-parity mixing that shatters the $Z_2$ symmetry, fundamentally reconstructing the KK profiles into states with extreme spatial polarization.
\end{itemize}

{The paper is organized as follows. Section~\ref{sec:unperturbed_basis} reviews the standard KK decomposition and establishes the unperturbed reference basis. Section~\ref{sec:mixing_mechanism} formalizes the mode mixing mechanism induced by dynamical bulk perturbations. Section~\ref{sec:physical_origins} investigates the specific physical origins of this mixing, analyzing both non-minimal dilaton couplings and geometric fluctuations. Finally, Section~\ref{sec:conclusion} summarizes our conclusions.}

\section{Kaluza-Klein Decomposition and the Unperturbed Reference Basis}
\label{sec:unperturbed_basis}

In a 5D braneworld model, we consider a free bulk fermion field coupled to a background scalar field $\phi$, described by the action
\begin{equation}\label{action_spinor}
    S_{F} = \int d^5x \, \sqrt{-G} \left[ \bar{\Psi} \Gamma^M ( \partial_M + \omega_M) \Psi - \eta \bar{\Psi} F(\phi) \Psi \right] ,
\end{equation}
where $G$ is the determinant of the 5D metric and $\omega_M$ denotes the spin connection. For a conformally flat metric parameterized as $ds^2 = e^{2A(z)} \big( \hat{g}_{\mu\nu}(x) dx^\mu dx^\nu + dz^2 \big)$, where $A(z)$ is the warp factor and $z$ is the extra-dimensional coordinate, the curved-space gamma matrices take the form $\Gamma^M = e^{-A} (\gamma^\mu, \gamma^5)$. Under this metric geometry, its non-vanishing components evaluate to $\omega_\mu = \frac{1}{2} (\partial_z A) \gamma_\mu \gamma_5 + \hat{\omega}_\mu$, while $\omega_z = 0$. 

Expanding the Dirac operator and rescaling the spinor field as $\Psi = e^{-2A} \hat{\Psi}$ to factor out the conformal weight, the action simplifies to
\begin{equation}\label{eq:L_mass_5D}
    S_{F} = \int d^4x dz \sqrt{-\hat{g}} \, \bar{\hat{\Psi}} \left[ \gamma^\mu(\partial_\mu + \hat{\omega}_\mu)+ \gamma^5 \partial_z - \eta e^A F(\phi) \right] \hat{\Psi} .
\end{equation}
To obtain the 4D effective action, we perform a chiral Kaluza-Klein (KK) decomposition
\begin{equation}\label{kkdecom}
    \hat{\Psi}(x,z) = \sum_n \left( \psi_{Ln}(x) \alpha_{Ln}(z) + \psi_{Rn}(x) \alpha_{Rn}(z) \right) ,
\end{equation}
and impose the standard orthonormalization conditions on the mode profiles, $\int dz \, \alpha_{Lm} \alpha_{Ln} = \int dz \, \alpha_{Rm} \alpha_{Rn} = \delta_{mn}$. Integrating out the extra dimension then yields the complete 4D effective action:
\begin{equation}\label{eq:standard_mass_term}
    S_{\text{eff}}^{(4D)} = \int d^4x \sqrt{-\hat{g}} \sum_{m,n} \left[ \delta_{mn} \bar{\psi}_{m} \gamma^\mu (\partial_\mu + \hat{\omega}_\mu)\psi_{n} - M_{mn}^{(LR)} \bar{\psi}_{Lm} \psi_{Rn} - M_{mn}^{(RL)} \bar{\psi}_{Rm} \psi_{Ln} \right] ,
\end{equation}
where the mass matrix elements are defined by
\begin{align}
    M_{mn}^{(LR)} &= \int_{-\infty}^{\infty} dz \, \alpha_{Lm}(z) \left[ -\partial_z + \eta e^{A(z)} F(\phi) \right] \alpha_{Rn}(z) ~, \label{eq:MLR} \\
    M_{mn}^{(RL)} &= \int_{-\infty}^{\infty} dz \, \alpha_{Rm}(z) \left[ \partial_z + \eta e^{A(z)} F(\phi) \right] \alpha_{Ln}(z) ~. \label{eq:MRL}
\end{align}
To ensure a Hermitian 4D effective Lagrangian, the mass matrices must satisfy $M^{(RL)} = (M^{(LR)})^\dagger$. Assuming real-valued background fields and KK profiles that vanish at the boundaries ($z \to \pm\infty$), integrating $M_{nm}^{(LR)}$ by parts directly yields
\begin{equation}
M_{nm}^{(LR)} = \int_{-\infty}^{\infty} dz , \alpha_{Rm}(z) \left[ \partial_z + \eta e^{A(z)} F(\phi) \right] \alpha_{Ln}(z) = M_{mn}^{(RL)} ~.
\end{equation}
This establishes the transpose relation $M^{(RL)} = (M^{(LR)})^T$, naturally guaranteeing the reality of the 4D mass terms.

To interpret the 4D effective theory as a tower of independent Dirac fermions, one must identify the true physical mass eigenstates that strictly diagonalize the mass matrix. In the idealized case of a bare, unperturbed background, we can define the reference KK profiles $\alpha_{L,Rn}(z)$ precisely to fulfill this requirement, which directly yields the familiar first-order coupled equations:
\begin{align}\label{eq:first_order}
    \left( \partial_z - \eta e^A F \right) \alpha_{Rn} &= -m_n \alpha_{Ln}, \nonumber \\
    \left( \partial_z + \eta e^A F \right) \alpha_{Ln} &= m_n \alpha_{Rn}.
\end{align}
By applying the corresponding adjoint operators, these first-order relations decouple into two independent second-order Schr\"{o}dinger-like equations:
\begin{align}\label{eq:schrodinger}
    \left[ -\partial_z^2 + V_L(z) \right] \alpha_{Ln} &= m_n^2 \alpha_{Ln}, \nonumber \\
    \left[ -\partial_z^2 + V_R(z) \right] \alpha_{Rn} &= m_n^2 \alpha_{Rn},
\end{align}
where $V_{L,R}(z) = \eta^2 e^{2A} F^2 \mp \partial_z(\eta e^A F)$.  {Crucially, the eigenvalues $m_n$ and eigenfunctions $\alpha_{L,Rn}(z)$ obtained from these decoupled equations should therefore be interpreted as the bare mass spectrum and an auxiliary orthogonal reference basis associated with the unperturbed background geometry, rather than the exact physical states of the full interacting theory.}

When additional background dynamics are introduced, one could, in principle, incorporate these operators directly into the equations of motion. However, because such extra terms typically spoil the simple symmetries of the unperturbed background, obtaining exact analytical solutions is generally a formidable task. Fortunately, since these additional dynamics usually manifest as small physical fluctuations, it is highly justified to treat them as perturbations. Instead of abandoning the analytical reference basis, we evaluate these perturbations within the unperturbed KK basis. 

This approach inevitably generates off-diagonal left-right couplings, leading to perturbative shifts in the physical mass spectrum and redefinitions of the true wavefunctions. To resolve these modified eigenstates, one must construct the perturbed mass matrix and perform a Singular Value Decomposition (SVD). 

\section{Mixing Mechanism: Dynamical Perturbations in the Bulk}
\label{sec:mixing_mechanism}

When the total extra-dimensional Dirac operator is shifted by a dynamical perturbation $\Delta \hat{D}$, the complete 5D action reads
\begin{equation}\label{eq:perturbed_action}
    S_{F}^{(\text{total})} = \int d^4x dz \sqrt{-\hat{g}} \, \bar{\hat{\Psi}} \left[ \gamma^\mu (\partial_\mu + \hat{\omega}_\mu) + \gamma^5 \partial_z - \eta e^{A(z)} F(\phi) + \Delta \hat{D} \right] \hat{\Psi} ~.
\end{equation}
Following the perturbative strategy outlined above, we directly expand the 5D spinor $\hat{\Psi}(x,z)$ in terms of the unperturbed chiral reference basis. Identifying the unperturbed bare differential operator as a purely scalar function $\hat{D}_0 \equiv -\partial_z + \eta e^{A(z)} F(\phi)$, and substituting the KK expansion back into Eq.~\eqref{eq:perturbed_action}, the 4D effective mass Lagrangian takes the form:
\begin{equation}\label{effetive4Dper}
    \mathcal{L}_{\text{mass}}^{(4D)} = - \sum_{m,n} \bar{\psi}_{Lm}(x) \left[ \int_{-\infty}^{\infty} dz \, \alpha_{Lm}^{(0)}(z) \left( \hat{D}_0 + \Delta D(z) \right) \alpha_{Rn}^{(0)}(z) \right] \psi_{Rn}(x) + \text{h.c.} .
\end{equation}

{Importantly, the unperturbed KK basis diagonalizes strictly the bare Dirac operator $\hat{D}_0$. Once a generic perturbation $\Delta \hat{D}$ is introduced, this original basis ceases to be the physical eigenbasis because the perturbation generally does not commute with the bare operator:
\begin{equation}
[\hat{D}_0, \Delta \hat{D}] \neq 0.
\end{equation}
Consequently, the original KK eigenstates can no longer define the exact physical particles of the interacting theory. Instead, the true physical spectrum must be reconstructed by diagonalizing the full interacting Dirac operator. In this sense, the interacting KK tower should be viewed as a collectively coupled spectral system rather than a mere collection of isolated Schr\"{o}dinger eigenmodes.}

By extracting the term in the square brackets \eqref{effetive4Dper}, we identify the exact matrix elements $M_{mn}$ of the new 4D effective mass matrix $M$. Because the reference basis is strictly tailored to diagonalize the bare operator $\hat{D}_0$, its action isolates the unperturbed diagonal mass matrix $m_n^{(0)} \delta_{mn}$. However, the perturbation operator $\Delta \hat{D}(z)$ shifts the total matrix elements, yielding
\begin{align}\label{eq:perturbed_matrix}
    M_{mn} &= \int_{-\infty}^{\infty} dz \, \alpha_{Lm}^{(0)} \left( \hat{D}_0 \right) \alpha_{Rn}^{(0)} + \int_{-\infty}^{\infty} dz \, \alpha_{Lm}^{(0)} \left( \Delta \hat{D} \right) \alpha_{Rn}^{(0)} \nonumber \\
    &= m_n^{(0)} \delta_{mn} + \Delta M_{mn} ~.
\end{align}

This formulation clearly illustrates the physical role of the perturbation. In realistic scenarios, $\Delta \hat{D}(z)$ possesses a non-trivial geometric profile along the extra dimension. Consequently, the overlap integral evaluated between two distinct orthogonal basis functions generally does not vanish ($\Delta M_{mn} \neq 0$ for $m \neq n$). The original KK modes are no longer independent physical states, but continuously transition into one another. To find the true physical particles and their exact masses, one must diagonalize this fully coupled mass matrix $M$, which yields the physical mass spectrum as the singular values of the perturbed system.

{
It is important to emphasize that the resulting Dirac mass matrix is generally non-Hermitian because the left- and right-handed sectors belong to distinct chiral spaces. The effective 4D mass term takes the form $\mathcal{L}_{\text{mass}}^{(4D)} = -\bar{\psi}_L M \psi_R + \text{h.c.} $, which cannot, in general, be diagonalized through a single unitary transformation. Instead, the exact physical mass spectrum must be obtained through a bi-unitary decomposition:
\begin{equation}\label{eq:svd_matrix}
    M = U_L \widetilde{M} U_R^\dagger ~,
\end{equation}
where $U_L$ and $U_R$ are two unitary matrices, and $\widetilde{M} = \text{diag}(\tilde{m}_0, \tilde{m}_1, \dots)$ is a diagonal matrix containing the real, non-negative singular values. This singular-value structure is the natural spectral characterization of the interacting Dirac operator, successfully resolving the fully coupled system into the true physical eigenstates.
}

{However, this mathematical non-Hermiticity of $M$ does not violate physical consistency. Because the underlying 5D differential operators strictly preserve the overall self-adjointness of the full Dirac system, the squared mass matrices $M M^\dagger$ and $M^\dagger M$ are mathematically guaranteed to be strictly Hermitian and positive semi-definite. Thus, this SVD framework inherently conserves probability and successfully resolves the fully coupled system into a mathematically consistent, orthonormal set of true physical mass eigenstates.}

The original 4D KK fields, $\psi_{Lm}$ and $\psi_{Rn}$, are related to the true physical mass eigenstates, denoted by $\tilde{\psi}_{Lk}$ and $\tilde{\psi}_{Rk}$, via the linear unitary transformations:
\begin{equation}\label{eq:field_redefinition}
    \psi_{Lm}(x) = \sum_{k} (U_L)_{mk} \tilde{\psi}_{Lk}(x) ~, \quad \psi_{Rn}(x) = \sum_{k} (U_R)_{nk} \tilde{\psi}_{Rk}(x) ~.
\end{equation}

Crucially, we must also verify that these transformations preserve the canonical structure of the kinetic terms. Substituting the field redefinitions into the unperturbed kinetic sector, and noting that the constant matrix elements commute with the 4D spacetime derivative $\partial_\mu$, we find for the left-handed sector:
\begin{align}
    \sum_m \bar{\psi}_{Lm} \gamma^\mu \partial_\mu \psi_{Lm} &= \sum_{m,j,k} \left( \bar{\tilde{\psi}}_{Lj} (U_L^\dagger)_{jm} \right) \gamma^\mu \partial_\mu \left( (U_L)_{mk} \tilde{\psi}_{Lk} \right) \nonumber \\
    &= \sum_{j,k} \bar{\tilde{\psi}}_{Lj} \gamma^\mu \partial_\mu \bigg(\sum_m (U_L^\dagger)_{jm} (U_L)_{mk} \bigg) \tilde{\psi}_{Lk} = \sum_k \bar{\tilde{\psi}}_{Lk} \gamma^\mu \partial_\mu \tilde{\psi}_{Lk} ~,
\end{align}
where we have used the unitarity condition $\sum_m (U_L^\dagger)_{jm} (U_L)_{mk} = \delta_{jk}$. An identical result holds for the right-handed sector $U_R$. Therefore, the kinetic terms remain perfectly diagonal and canonically normalized. Combining this invariant kinetic sector with the diagonalized mass sector yields the complete physical 4D effective Lagrangian:
\begin{align}\label{eq:physical_lagrangian}
    \mathcal{L}_{\text{eff}}^{(4D)} &= \sum_k \left[ \bar{\tilde{\psi}}_{Lk} \gamma^\mu \partial_\mu \tilde{\psi}_{Lk} + \bar{\tilde{\psi}}_{Rk} \gamma^\mu \partial_\mu \tilde{\psi}_{Rk} - \tilde{m}_k \left( \bar{\tilde{\psi}}_{Lk} \tilde{\psi}_{Rk} + \bar{\tilde{\psi}}_{Rk} \tilde{\psi}_{Lk} \right) \right] \nonumber \\
    &= \sum_{k} \bar{\tilde{\psi}}_{k} \left( \gamma^\mu \partial_\mu - \tilde{m}_k \right) \tilde{\psi}_{k} ~.
\end{align}
This final expression elegantly recovers the standard 4D effective action for a tower of independent Dirac fermions. The singular values $\tilde{m}_k$ rigorously define the true physical masses of the $k$-th fermion KK modes, incorporating all dynamical mode mixing effects induced by the bulk perturbations.

Beyond merely shifting the mass eigenvalues to the singular values $\tilde{m}_k$, this unitary mixing inherently reconstructs the extra-dimensional wavefunctions of the physical states. To maintain consistency with the field redefinitions under the KK decomposition, the exact physical profiles along the extra dimension must transform via the adjoint matrices, given explicitly by:
\begin{equation}\label{eq:physical_profiles}
    \tilde{\alpha}_{Lk}(z) = \sum_m (U_L^\dagger)_{km} \alpha_{Lm}^{(0)}(z) ~, \quad
    \tilde{\alpha}_{Rk}(z) = \sum_n (U_R^\dagger)_{kn} \alpha_{Rn}^{(0)}(z) ~.
\end{equation}
Consequently, the probability density distributions of the physical fermions along the extra dimension are deformed. This spatial reshaping is of profound phenomenological significance: it directly modifies the overlap integrals with other bulk fields or brane-localized operators, thereby dynamically altering the effective 4D gauge and Yukawa couplings of the physical KK tower.

\section{Physical Origins of Bulk Perturbations}\label{sec:physical_origins}
We outline two physically well-motivated scenarios where such perturbative operators naturally arise.

\subsection{Non-Minimal Dilaton-Fermion Couplings} \label{dilatoncouplings}

From an effective field theory perspective, treating the dilaton interaction as a dynamical perturbation is rigorously justified by an intrinsic hierarchy of energy scales. In a realistic two-field thick brane scenario \cite{Guo2014fba}, the topological kink field $\phi(z)$ is the primary source generating the core brane tension. Its coupling provides the dominant confining potential well that traps the fermion zero mode. Conversely, the dilaton field $\pi(z)$ typically governs subdominant background dynamics. To prevent the dilaton's gravitational backreaction from destabilizing the primary topological defect, its energy density and effective interaction scale must be parametrically suppressed. Parameterizing the generic additive couplings as $\eta \bar{\Psi}F(\phi) \Psi + \xi \bar{\Psi} \text{e}^{\lambda\pi} \Psi$ directly reflects this physical scale separation: $|\xi \text{e}^{\lambda\pi}| \ll |\eta F(\phi)|$ within the localization region. This hierarchy allows us to cleanly decouple the background dynamics. The dominant scalar interaction is entirely absorbed into the bare operator $\hat{D}_0$, establishing the exact unperturbed reference basis $\{ \alpha_{L,Rn}^{(0)} \}$. The subdominant dilaton coupling is then cleanly isolated as the perturbation operator:
\begin{equation}\label{perb1}
    \Delta \hat{D} = \xi \text{e}^{\lambda\pi} ~.
\end{equation} 
This formulation not only ensures theoretical stability but also isolates the exact geometric contribution of the dilaton field responsible for driving the KK mode mixing.

To illustrate this quantitatively, we evaluate this framework numerically. We consider the two-field thick brane model introduced in Ref.~\cite{Fu2011}, which admits the following analytical background solutions:
\begin{subequations}\label{branesol}
\begin{eqnarray}
    \phi(z) &=& v \tanh(a z) ~, \label{eq:phi} \\
    A(z) &=& -\frac{v^2}{9} \big( \ln \cosh^2(a z) + \frac{1}{2} \tanh^2(a z) \big) ~, \label{eq:A} \\
    \pi(z) &=& \sqrt{3} \, A(z) ~, \label{eq:pi}
\end{eqnarray}
\end{subequations}
where $\pi(z)$ represents the dilaton field, and $v, a$ are constants. Following the coupling structure introduced above, we specify the dominant scalar interaction as $F(\phi) = \text{e}^{-A(z)}\phi(z)$. In this setup, the unperturbed effective potential is entirely dictated by the kink field $\phi$, while the dilaton field strictly serves as the source of the explicit perturbation \eqref{perb1}.

The unperturbed effective potentials exhibit a Pöschl-Teller-like structure, which naturally supports a discrete spectrum of localized bound fermion KK modes. By first solving for these unperturbed basis states, we explicitly evaluate the perturbed mass matrix $M_{mn}$ for various values of the Yukawa coupling $\eta$. The resulting matrices and the corresponding physical mass spectra $\tilde{m}_k$ are presented in table~\ref{case1massmatrix}, where we have fixed the model parameters to $\xi=0.2, \lambda=0.05$, $v=1.0$, and $a=1.0$.

\begin{table*}[htbp]
\centering
\caption{Unperturbed KK masses ($m_n$), perturbed mass mixing matrices ($M_{nm}$), the resulting physical mass spectra ($\tilde{m}_k$), and relative mass shifts ($\delta m/m$) for varying Yukawa couplings $\eta$.}
\label{case1massmatrix}
\footnotesize 
\setlength{\tabcolsep}{8pt}    
\setlength{\arraycolsep}{3pt}   
\begin{NiceTabular}{c|c|c|c|c}
\Hline
& $\eta=4$ & $\eta=3$ & $\eta=2$ & $\eta=1.5$ \\
\Hline
\multirow{5}{*}{$m_n$ } 
& 0 / - & 0 / - & 0 / - & 0 / - \\
& 2.65  & 2.24  & 1.73  & 1.42  \\
& 3.45  & 2.83  & 2.05  & -     \\
& 3.87  & 3.04  & -     & -     \\
& 4.03  & -     & -     & -     \\
\Hline
$\{M_{nm}\}$ 
& $\begin{bmatrix}
0.99 & 0.00 & -0.09 & 0.00 \\
2.65 & 0.97 & 0.00 & -0.14 \\
0.10 & 3.46 & 0.89 & 0.00 \\
0.00 & 0.21 & 3.87 & 0.80 \\
-0.01 & 0.00 & 0.36 & 4.03
\end{bmatrix}$ 
& $\begin{bmatrix}
0.20 & 0.00 & -0.02 \\
2.24 & 0.18 & 0.00 \\
0.03 & 2.83 & 0.16 \\
0.00 & 0.06 & 3.03
\end{bmatrix}$ 
& $\begin{bmatrix}
0.19 & 0.00 \\
1.73 & 0.17 \\
0.05 & 2.04
\end{bmatrix}$ 
& $\begin{bmatrix}
0.18 \\
1.42 
\end{bmatrix}$ \\
\Hline
 $ \tilde{m}_k$ 
& (0, 2.64, 3.45, 3.84, 4.10) 
& (0, 2.23, 2.80, 3.09) 
& (0, 1.71, 2.07) 
& (0, 1.43) \\
\Hline
 $\frac{\delta m}{m}$ (\%)
& \tiny (-, -0.4\%, 0.0\%, -0.8\%, +1.7\%) 
& \tiny(-, -0.4\%, -1.1\%, +1.6\%) 
& \tiny(-, -1.2\%, +1.0\%) 
& \tiny(-, +0.7\%) \\
\Hline
\end{NiceTabular}
\end{table*}

As clearly demonstrated by the numerical results in table~\ref{case1massmatrix}, the physical mass spectra $\tilde{m}_k$ undergo substantial corrections compared to the unperturbed eigenvalues $m_n$. A systematic analysis of these shifts reveals several underlying regularities driven by the non-trivial SVD mixing:
\begin{itemize}
    \item \textbf{Zero-mode protection:} The lowest physical mass $\tilde{m}_0$ remains strictly zero. This confirms that the 5D chirality of the topological zero mode is robustly protected and cannot acquire a mass purely through bulk mode mixing.
    \item \textbf{Quantum level repulsion:} As explicitly shown in Table~\ref{case1massmatrix}, the off-diagonal geometric couplings induce a classic quantum mechanical level repulsion effect. According to second-order perturbation theory, the absolute mass shift of the $n$-th state, defined as {$\Delta(m_n^2) \approx \sum_{k \neq n} \frac{|\langle n | \delta(M^\dagger M) | k \rangle|^2}{m_n^2 - m_k^2}$}, is theoretically approximated by $\Delta m_n \approx \sum_{k \neq n} |M_{nk}|^2 / (m_n - m_k)$. Because the lowest massive excitation ($m_1$) only interacts with higher-lying states ($m_1 - m_k < 0$), it consistently experiences a net downward pressure, resulting in a systemic mass reduction. Conversely, the uppermost bound states within the finite potential well are repelled upwards by the underlying lower modes. To quantitatively evaluate the phenomenological impact of this repulsion, we introduce the relative mass shift rate, defined as $\delta m / m \equiv \Delta m_n / m_n$. As presented in the table, in the regime supporting multiple bound states (e.g., $\eta=4$), this mechanism yields a $-0.4\%$ downward correction for the first excited state ($m_1$) and a strong $+1.7\%$ upward repulsion for the uppermost state ($m_4$), cleanly demonstrating the spectral redistribution under the dilaton perturbation. 

\end{itemize}

\begin{figure}[htbp]
    \centering
    \includegraphics[width=\textwidth]{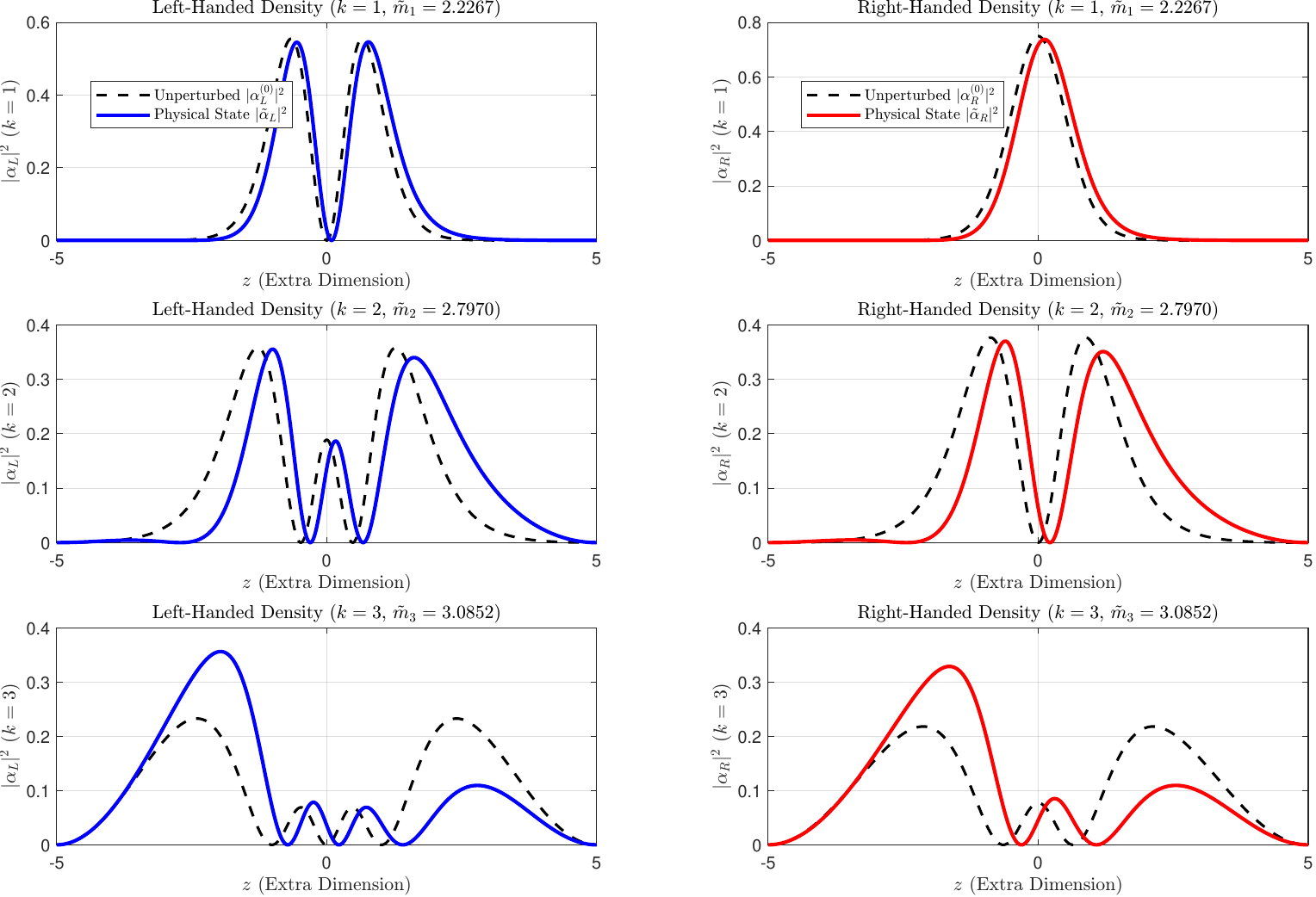}
    \caption{The probability density distributions of the unperturbed (dashed lines) and physical (solid lines) KK states for $\eta = 2.00$ and $\lambda = 0.05$.}
    \label{case1density}
\end{figure}

Beyond the shifts in the mass spectrum, this dynamically induced mixing inherently reconstructs the extra-dimensional wavefunctions of the physical states. To visually comprehend this effect, we present the probability density distributions of the physical states in figure~\ref{case1density} for $\eta = 2$.
The most striking feature observed in Figure~\ref{case1density} is the severe geometric asymmetry of the physical KK modes. 

This parity-breaking deformation is mathematically rooted in the structure of the perturbed mass matrix. Because the background dilaton perturbation \eqref{perb1} is a parity-even geometric fluctuation, its overlap integral exclusively couples left- and right-handed basis states that share the same parity. However, because the full mass matrix $M_{mn}$ superimposes the unperturbed diagonal masses with the perturbative shifts, it completely loses any definitive block-diagonal or checkered structure. Consequently, the left- and right-handed sectors can no longer be decoupled into independent parity channels, forcing the SVD unitary matrices $U_L$ and $U_R$ to mix parity-even and parity-odd unperturbed modes simultaneously. This structural scrambling inevitably shatters the geometric reflection symmetry of the physical wavefunctions, which dynamically manifests as the spatial polarization observed. 


The physical significance of this spatial asymmetry is profound for braneworld phenomenology. Because the 4D effective coupling strength of these fermions to Standard Model fields localized on the brane ($z=0$) is dictated by their probability density at the origin, this geometric shift dynamically redefines their interaction profiles. Most notably, unperturbed states that originally possessed a geometric node at the brane (e.g., the left-handed $k=1$ and right-handed $k=2$ states) acquire a substantial, non-zero probability density at the origin due to this asymmetric mixing. Consequently, the background perturbation effectively `illuminates' these previously dark KK modes, generating non-vanishing effective couplings and opening novel channels for their potential observation in 4D colliders.

\subsection{Geometric Fluctuations} 
\label{sec:geometric_fluctuations}

In realistic braneworld scenarios, the background spacetime geometry is dynamically determined by the bulk energy-momentum tensor. Any dynamical perturbations in the bulk—such as radion excitations, phase transitions, or the vacuum expectation values of bulk fields—naturally carry an energy density that sources a physical gravitational backreaction. Through the linearized Einstein equations, this backreaction statically deforms the background warp factor, leading to a metric fluctuation $A(z) = A_0(z) + \delta A(z)$. 

For bulk fermions, the warp factor is intimately tied to both the vielbein and the spin connection. Consequently, this metric fluctuation cascades directly into the 5D Dirac operator. By expanding the kinetic and Yukawa terms up to the linear order in $\delta A(z)$, the effective shift in the Dirac operator is given by
\begin{equation}\label{DeltaD}
    \Delta \hat{D}_{\text{geom}} \simeq \left( \delta A(z) \right) \partial_z + \eta e^{A_0(z)} \delta A(z) F(\phi) ~.
\end{equation}
Acting as a $z$-dependent perturbative operator, this operator introduces off-diagonal couplings within the unperturbed KK representation, inevitably driving dynamical mixing among the previously independent orthogonal KK modes. In the following, we systematically investigate the mode mixing and wavefunction deformations induced by both symmetric and asymmetric geometric fluctuations.

\subsubsection{Even-Parity Geometric Perturbation}

Since the energy density of the bulk fields is typically localized symmetrically around the brane core, it is phenomenologically well-motivated to model the induced gravitational backreaction as a strictly even-parity Gaussian deformation:
\begin{equation}
    \delta A(z) = \epsilon e^{-\alpha z^2} ~,
\end{equation}
where $\epsilon$ is a dimensionless small parameter characterizing the amplitude of the geometric fluctuation, and $\alpha$ determines its effective width. This explicitly even-parity form, mathematically satisfying $\delta A(-z) = \delta A(z)$, correctly captures the localized nature of the metric distortion while safely vanishing at $|z| \to \infty$. This ensures the preservation of the asymptotic spacetime geometry and guarantees the absolute convergence of the overlap integrals.

Substituting this into the general perturbation operator yields the specific off-diagonal mass matrix elements generated:
\begin{equation}
    \delta M_{mn}^{(\text{geom})} = \epsilon \int_{-\infty}^{\infty} dz \, e^{-\alpha z^2} \alpha_{L,m}^{(0)}(z) \left[ \partial_z + \eta e^{A_0(z)} F(\phi) \right] \alpha_{R,n}^{(0)}(z) ~.
\end{equation}
Diagonalizing the total mass matrix, $M_{mn} = m_n^{(0)}\delta_{mn} + \delta M_{mn}^{(\text{geom})}$, yields the physical mass spectrum modified by the gravitational backreaction.

\begin{figure}[htbp]
    \centering
    \includegraphics[width=\textwidth]{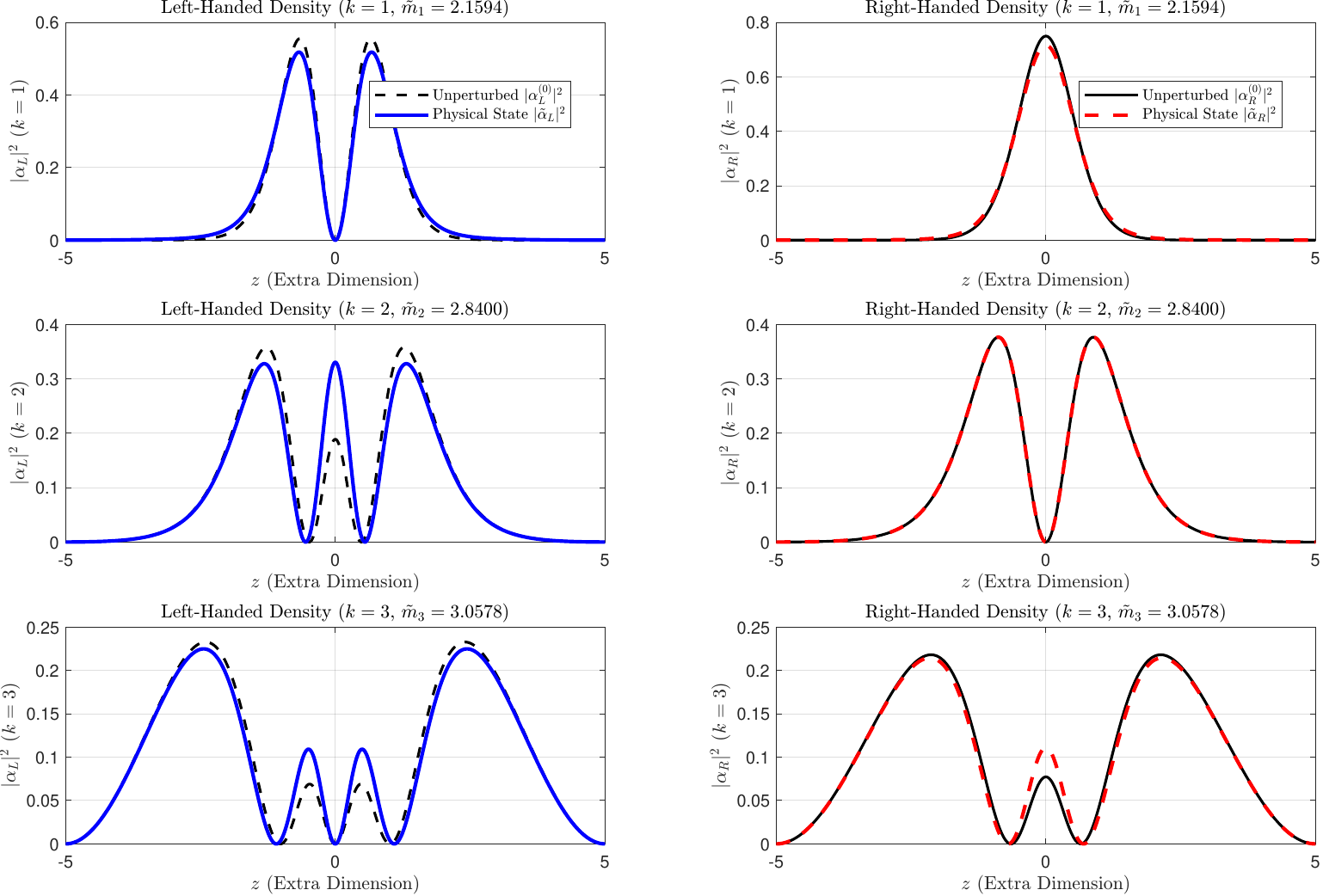}
    \caption{The probability density distributions of the unperturbed (dashed lines) and physical (solid lines) KK states under even-parity geometric fluctuations ($\eta=3.0$, $\epsilon=0.3$, $\alpha=1.0$). }
    \label{case2density}
\end{figure}

To provide concrete numerical results, we evaluate the specific 5D thick brane background solution given in Eq.~\eqref{branesol}, utilizing the scalar coupling $F(\phi) = e^{-A_0(z)}\phi(z)$. The explicitly calculated mass matrices and the resulting physical mass spectra for different values of $\eta$ are listed in table~\ref{case2massmatrix}, alongside the corresponding reconstructed probability density distributions illustrated in figure~\ref{case2density}. For these numerical evaluations, the background brane solution is fixed by $v=1.0$ and $a=1.0$, with the geometric perturbation parameters chosen as $\epsilon=0.3$ and $\alpha=1.0$.

\begin{table*}[htbp]
\centering
\caption{Unperturbed KK masses ($m_n$), perturbed mass mixing matrices ($M_{nm}$), the resulting physical mass spectra ($\tilde{m}_k$), and relative mass shifts ($\delta m/m$) for varying Yukawa couplings $\eta$ under the even-parity geometric fluctuation scenario.}
\label{case2massmatrix}
\footnotesize 
\setlength{\tabcolsep}{8pt}    
\begin{NiceTabular}{c|c|c|c|c}
\Hline
& $\eta=4$ & $\eta=3$ & $\eta=2$ & $\eta=1.5$ \\
\Hline
\multirow{5}{*}{$m_n$ } 
& 0 / - & 0 / - & 0 / - & 0 / - \\
& 2.65  & 2.24  & 1.73  & 1.42  \\
& 3.45  & 2.83  & 2.05  & -     \\
& 3.87  & 3.04  & -     & -     \\
& 4.03  & -     & -     & -     \\
\Hline
$\{M_{nm}\}$ 
& {\scriptsize $\left[ \begin{matrix}
 0.00 & -0.61 & -0.00 &  0.19 \\
 2.57 &  0.00 & -0.41 &  0.00 \\
 0.00 &  3.42 &  0.00 & -0.13 \\
 0.02 &  0.00 &  3.88 & -0.00 \\
 0.00 &  0.02 & -0.00 &  4.05
\end{matrix} \right]$} 
& {\scriptsize $\left[ \begin{matrix}
-0.00 & -0.43 &  0.00 \\
 2.16 &  0.00 & -0.16 \\
 0.00 &  2.81 & -0.00 \\
 0.02 & -0.00 &  3.05
\end{matrix} \right]$} 
& {\scriptsize $\left[ \begin{matrix}
 0.00 & -0.20 \\
 1.67 &  0.00 \\
-0.00 &  2.05
\end{matrix} \right]$} 
& {\scriptsize $\left[ \begin{matrix}
0.00 \\
1.37 
\end{matrix} \right]$} \\
\Hline
 $ \tilde{m}_k$ 
& (0, 2.55, 3.47, 3.92, 4.06) 
& (0, 2.16, 2.84, 3.06) 
& (0, 1.67, 2.06) 
& (0, 1.37) \\
\Hline
 $\frac{\delta m}{m}$ (\%)
& \tiny (-, -3.8\%, +0.6\%, +1.3\%, +0.7\%) 
& \tiny (-, -3.6\%, +0.4\%, +0.7\%) 
& \tiny (-, -3.5\%, +0.5\%) 
& \tiny (-, -3.5\%) \\
\Hline
\end{NiceTabular}
\end{table*}

Since the Gaussian fluctuation $e^{-\alpha z^2}$ is even and the unperturbed operator $[\partial_z \mp \eta e^{A_0(z)} F(\phi)]$ is mathematically odd, the overall perturbation $\Delta \hat{D}_{\text{geom}}$ is strictly an odd function of $z$. Consequently, a non-vanishing overlap integral requires the left- and right-handed profiles to possess opposite spatial parities. Given the alternating parities of the unperturbed KK basis, this rule strictly forbids nearest-neighbor transitions (e.g., $M_{01} = 0$) and exclusively couples states of the \textit{same} macroscopic parity (e.g., $M_{02} \neq 0$), yielding the mathematically decoupled checkerboard pattern explicitly observed in Table~\ref{case2massmatrix}. This strict selection rule naturally drives two profound physical consequences. First, by only connecting states separated by at least two KK levels, the enlarged unperturbed mass gaps strongly suppress the off-diagonal mixing, ensuring the relative mass shifts ($\delta m/m$) remain perturbatively small. Second, because the SVD exclusively superimposes basis states of identical $Z_2$ parity, the physical eigenstates perfectly preserve their macroscopic reflection symmetries. As vividly illustrated in Figure~\ref{case2density}, the density profiles exhibit no spatial polarization; instead, the symmetric backreaction exclusively induces internal amplitude `squeezing' or `stretching', gracefully retaining the original geometric balance.

Phenomenologically, this implies that symmetric gravitational backreactions cannot `illuminate' dark modes. Parity-odd physical states (e.g., the left-handed $k=1$ and right-handed $k=2$ modes) retain their exact geometric nodes at $z=0$, ensuring they remain completely decoupled from 4D brane-localized fields. Meanwhile, the effective couplings of parity-even states (e.g., the left-handed $k=2$ and right-handed $k=1$ modes, which exhibit wave antinodes at $z=0$) are simply rescaled by the symmetric amplitude deformations, preserving their fundamental interaction channels.

\subsubsection{Odd-Parity Geometric Perturbation}

While symmetric geometric fluctuations provide a standard baseline for studying mode mixing, realistic braneworld scenarios may involve asymmetric dynamics. Discrepancies between the bulk vacuum states on either side of the brane, phase transitions, or asymmetric matter condensates can source an asymmetric energy-momentum tensor. Through the linearized Einstein equations, this induces a parity-breaking gravitational backreaction on the warp factor, $A(z) = A_0(z) + \delta A(z)$. 

To rigorously capture this phenomenon while ensuring mathematical consistency, we model the asymmetric geometric fluctuation using an odd-parity core modulated by a Gaussian envelope:
\begin{equation} \label{eq:asym_deltaA}
    \delta A(z) = \epsilon \tanh(\beta z) e^{-\gamma z^2}.
\end{equation}
The physical parameters in this ansatz are defined as follows:
\begin{itemize}
    \item $\epsilon$ is a dimensionless small parameter representing the overall amplitude of the asymmetric gravitational backreaction.
    \item $\beta$ determines the steepness of the asymmetry near the brane core ($z=0$), reflecting the gradient of the underlying asymmetric source.
    \item $\gamma$ characterizes the inverse squared width of the localized deformation. The Gaussian factor $e^{-\gamma z^2}$ is crucial as it guarantees that the metric perturbation safely vanishes at spatial infinity ($|z| \to \infty$), thereby preserving the asymptotic behavior of the background spacetime and ensuring the absolute convergence of the overlap integrals.
\end{itemize}

Substituting Eq.~(\ref{eq:asym_deltaA}) into the perturbed Dirac operator, and utilizing the specific scalar coupling $F(\varphi) = e^{-A_0(z)}\phi(z)$, the resulting off-diagonal mass matrix element takes the following elegant form:
\begin{equation} \label{eq:mass_matrix_asym}
    \delta M_{mn}^{(asym)} = \epsilon \int_{-\infty}^{\infty} dz \, \tanh(\beta z) e^{-\gamma z^2} \alpha_{Lm}^{(0)}(z) \left[ \partial_z + \eta \phi(z) \right] \alpha_{Rn}^{(0)}(z).
\end{equation}

\begin{figure}[htbp]
    \centering
    \includegraphics[width=\textwidth]{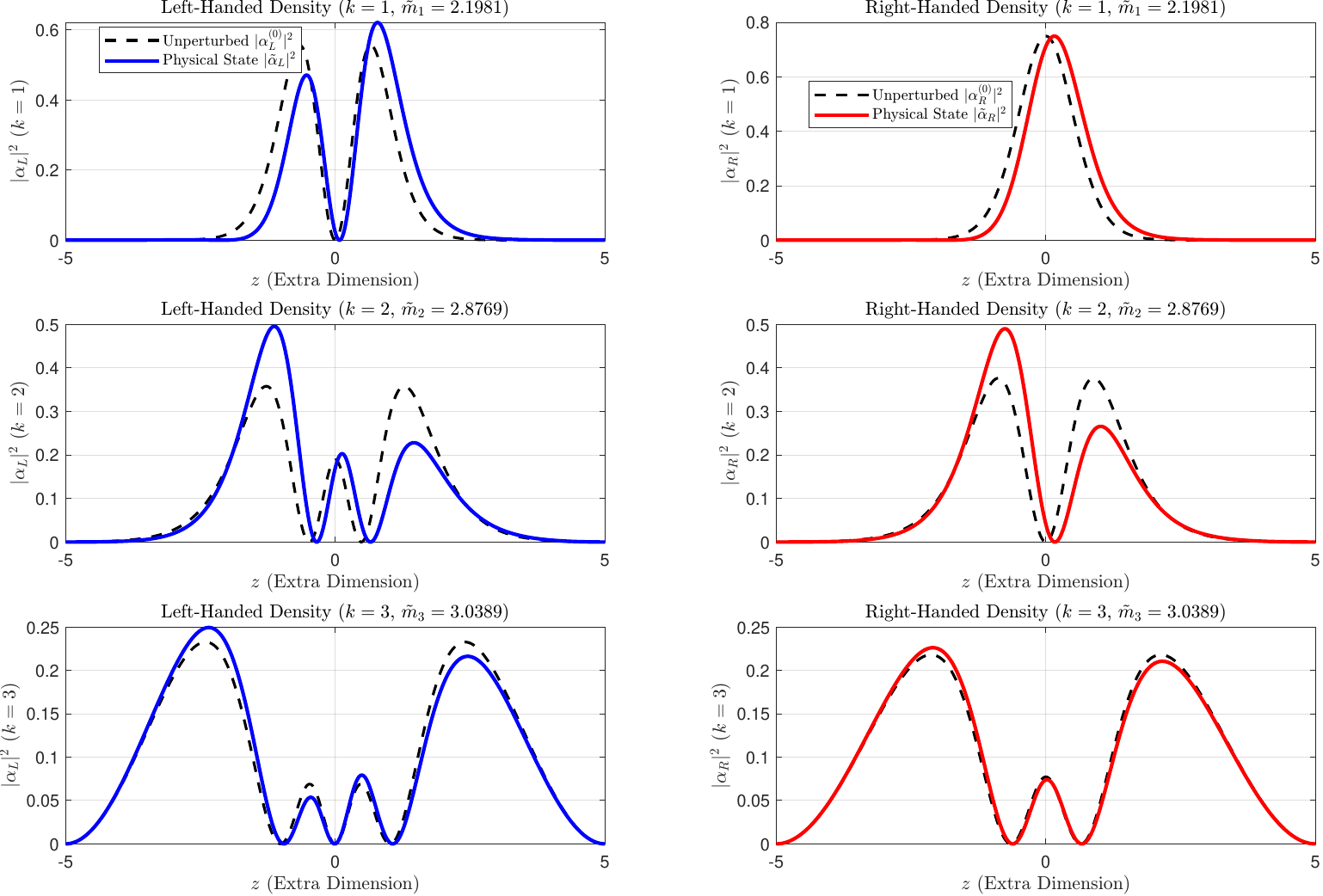}
    \caption{The probability density distributions of the unperturbed (dashed lines) and physical (solid lines) KK states under odd-parity geometric fluctuations ($\eta=3.0$, $\epsilon=0.3$, $\beta=4.0$, $\gamma=1.0$).}
    \label{case3density}
\end{figure}

To provide concrete numerical results, we evaluate the specific 5D thick brane background solution given in Eq.~\eqref{branesol}. The explicitly calculated mass matrices and the resulting physical mass spectra for different values of $\eta$ are listed in table~\ref{case3massmatrix}, and the physical wavefunctions, as visually presented in figure~\ref{case3density} for $\eta=3.0$. For these numerical evaluations, the background brane solution is fixed by $v=1.0$ and $a=1.0$, with the geometric perturbation parameters chosen as $\epsilon=0.3$, $\beta=4.0$, and $\gamma=1.0$.

\begin{table*}[htbp]
\centering
\caption{Unperturbed KK masses ($m_n$), perturbed mass mixing matrices ($M_{nm}$), the resulting physical mass spectra ($\tilde{m}_k$), and relative mass shifts ($\delta m/m$) for varying Yukawa couplings $\eta$ under the odd-parity geometric fluctuation scenario.}
\label{case3massmatrix}
\footnotesize 
\setlength{\tabcolsep}{8pt}    
\begin{NiceTabular}{c|c|c|c|c}
\Hline
& $\eta=4$ & $\eta=3$ & $\eta=2$ & $\eta=1.5$ \\
\Hline
\multirow{5}{*}{$m_n$ } 
& 0 / - & 0 / - & 0 / - & 0 / - \\
& 2.65  & 2.24  & 1.73  & 1.42  \\
& 3.45  & 2.83  & 2.05  & -     \\
& 3.87  & 3.04  & -     & -     \\
& 4.03  & -     & -     & -     \\
\Hline
$\{M_{nm}\}$ 
& {\scriptsize $\left[ \begin{matrix}
 0.06 &  0.00 &  0.27 & -0.00 \\
 2.65 &  0.44 & -0.00 & -0.04 \\
 0.03 &  3.46 &  0.25 & -0.00 \\
-0.00 & -0.11 &  3.87 &  0.10 \\
-0.01 &  0.00 & -0.07 &  4.03
\end{matrix} \right]$} 
& {\scriptsize $\left[ \begin{matrix}
 0.06 &  0.00 &  0.11 \\
 2.24 &  0.31 & -0.00 \\
 0.02 &  2.83 &  0.09 \\
 0.00 & -0.07 &  3.04
\end{matrix} \right]$} 
& {\scriptsize $\left[ \begin{matrix}
 0.06 & -0.00 \\
 1.73 &  0.14 \\
 0.00 &  2.04
\end{matrix} \right]$} 
& {\scriptsize $\left[ \begin{matrix}
0.06 \\
1.42 
\end{matrix} \right]$} \\
\Hline
 $ \tilde{m}_k$ 
& (0, 2.60, 3.52, 3.90, 4.03) 
& (0, 2.20, 2.88, 3.04) 
& (0, 1.72, 2.06) 
& (0, 1.42) \\
\Hline
 $\frac{\delta m}{m}$ (\%)
& \tiny (-, -1.9\%, +2.0\%, +0.8\%, 0.0\%) 
& \tiny (-, -1.8\%, +1.8\%, 0.0\%) 
& \tiny (-, -0.6\%, +0.5\%) 
& \tiny (-, 0.0\%) \\
\Hline
\end{NiceTabular}
\end{table*}

As explicitly demonstrated in Figure~\ref{case3density}, the dynamically forced nearest-neighbor mixing yields severe spatial polarization. This is because although the geometric perturbation \eqref{eq:asym_deltaA} possesses odd parity, the entire integrand defining the perturbation matrix elements \eqref{eq:mass_matrix_asym} is parity-even, analogous to the dilaton case. Notably, this symmetry-breaking deformation is visually and numerically concentrated within intermediate-mass states (e.g., $k=1, 2$), leaving highly excited states (e.g., $k=3$) remarkably immune.

This confinement originates directly from a spatial topological matching mechanism. The geometric source $\delta A(z) \propto \tanh(\beta z) e^{-\gamma z^2}$ generates a localized, double-peaked off-center topography with a strict node at the brane core ($z=0$). This specific topography acts as a spatial filter, perfectly aligning with the antinodes of intermediate KK states to trigger a ``lock-and-key'' spatial resonance that maximizes their mixing matrix elements. Consequently, the resulting mass corrections ($\delta m/m$) reach their maximum precisely within these intermediate-mass states (e.g., $k=1, 2$), which in turn drives a correspondingly severe deformation in their physical wavefunctions. In stark contrast, in the dilaton scenario, the largest mass corrections occur at the high-mass tail of the spectrum, consequently precipitating the most drastic adjustments to their wavefunctions.

\section{Conclusion}
\label{sec:conclusion}

{In this work, we have systematically investigated the mode mixing of fermionic Kaluza-Klein states induced by generic dynamical background perturbations. To rigorously evaluate the effects of these perturbations, we utilized the solutions of the isolated, unperturbed Schr"{o}dinger-like equations as our fundamental bare basis. Expanding the fully interacting Dirac operator within this reference basis inevitably generates off-diagonal couplings, resulting in a general, non-Hermitian 4D effective mass matrix. To completely resolve this fully coupled system and determine the true physical spectrum and wavefunctions, we employed a bi-unitary Singular Value Decomposition (SVD). This mathematical framework naturally accommodates the chiral asymmetry of the matrix while preserving the overall probability, allowing us to exact the physical mass eigenvalues and visualize the geometric reshaping of the corresponding states.}

Through systematic analysis of background perturbations---including both dilaton couplings and geometric backreactions---we uncovered that the intrinsic geometric reshaping of the physical wavefunctions is strictly dictated by the algebraic parity of the total perturbation operators. Specifically, parity-odd perturbation operators yield block-diagonal mass matrices, driving pure same-parity state mixing that perfectly preserves the macroscopic $Z_2$ spatial symmetry via amplitude modulations. In stark contrast, parity-even perturbation operators trigger severe cross-parity mixing that shatters the spatial $Z_2$ symmetry, fundamentally reconstructing the KK profiles into states with extreme spatial polarization along the extra dimension.

{Beyond the formal theoretical framework, this dynamical spatial reshaping carries profound phenomenological implications. Specifically, the parity-breaking polarization severely alters the wave function profiles, inducing non-vanishing probability densities at the brane location ($z=0$) for states that are strictly decoupled in the unperturbed limit. This geometric redefinition of overlap integrals naturally illuminates'' these previously dark'' KK modes, providing a robust theoretical mechanism for their emergence and offering new avenues for future phenomenological exploration.}


\providecommand{\href}[2]{#2}\begingroup\raggedright\endgroup

\end{document}